\documentclass[twocolumn]{svjour3}
\usepackage{graphics}
\usepackage{amsmath}
\usepackage{amssymb}

\journalname{epja}
\usepackage[utf8]{inputenc}
\usepackage[T1]{fontenc}

\usepackage{mathtools}
\usepackage{amssymb}
\usepackage{bm}
\usepackage{siunitx}
\usepackage{booktabs}
\usepackage{csquotes}

\DeclareMathOperator{\arsinh}{arsinh}\renewcommand*{\arsinh}{\sinh^{-1}}
\DeclareMathOperator{\arcosh}{arcosh}\renewcommand*{\arcosh}{\cosh^{-1}}
\DeclareMathOperator{\artanh}{artanh}\renewcommand*{\artanh}{\tanh^{-1}}

\DeclarePairedDelimiter{\parens}{\lparen}{\rparen}
\DeclarePairedDelimiter{\bracks}{\lbrack}{\rbrack}

\newcommand*{\x}{\mathclose{}\mathopen{\mskip\thinmuskip}}
\newcommand*{\diff}[1]{\mathrm{d}#1}
\newcommand*{\derivop}[1]{\partial_{#1}}
\newcommand*{\deriv}[2]{\frac{\mathopen{\partial}#1}{\mathopen{\partial}#2}}

\NewDocumentCommand{\tup}{smo}{\IfNoValueTF{#3}{\bm{#2}}{\IfBooleanTF{#1}{{#2}_{#3}}{{#2}^{#3}}}}
\DeclarePairedDelimiterX{\tupdelims}[1]{\lparen}{\rparen}{\cramped{#1}}


\DeclarePairedDelimiterX{\setdelims}[1]{\lbrace}{\rbrace}{\cramped{#1}}

\DeclarePairedDelimiterXPP{\func}[2]{#1}{\lparen}{\rparen}{}{\cramped{#2}}

\NewDocumentCommand{\drift}{om}{\tup{\mu}[#1]_{#2}}
\NewDocumentCommand{\diffusion}{om}{\tup{\sigma}[#1]_{#2}}
\NewDocumentCommand{\diffusivity}{om}{\tup{D}[#1]_{#2}}
\newcommand*{\pdf}[1]{f_{#1}}
\newcommand*{\fpo}[1]{L_{#1}}
\newcommand*{\potential}[1]{\Phi_{#1}}

\newcommand*{\equ}{\mathrm{equ}}

\NewDocumentCommand{\rap}{o}{\tup{\xi}[#1]}
\NewDocumentCommand{\Rap}{o}{\tup{\Xi}[#1]}

\usepackage{hyperref}

\begin{document}
%
\title{Spectral eigenfunction decomposition of a Fokker--Planck operator 
for relativistic heavy-ion collisions}
\author{A.\,Rizzi \and G.\,Wolschin}
\institute{Institut f\"ur Theoretische Physik der Universit\"at Heidelberg, Philosophenweg 16, D-69120 Heidelberg, Germany, EU}
\date{Received: date / Revised version: date}
%
\abstract{
A spectral solution method is proposed to solve a previously developed non-equilibrium statistical model describing partial thermalization of produced charged hadrons in relativistic heavy-ion collisions, thus improving the accuracy of the numerical solution. The particle’s phase-space trajectories are treated as a drift-diffusion stochastic process, leading to a Fokker--Planck equation (FPE) for the single-particle probability distribution function. The drift and diffusion coefficients are derived from the expected asymptotic states via appropriate fluctuation--dissipation relations, and the resulting FPE is then solved numerically using a spectral eigenfunction decomposition. The calculated time-dependent particle distributions are compared to Pb-Pb data from the ATLAS and ALICE collaborations at the Large Hadron Collider.
} 
\maketitle
\section{Introduction}
\label{intro}
Relativistic heavy-ion collisions  \cite{hot23} are a versatile tool to study the
partial approach of a quantum many-body system towards
statistical equilibrium. To account for time-dependent charged-hadron production at energies reached at the Relativistic Heavy Ion Collider RHIC and
the Large Hadron Collider LHC,  
an explicit and rigorous derivation of the Fokker--Planck equation (FPE) for the momentum distribution function in
longitudinal and transverse rapidities has recently been given \cite{hgw24}. Using the mathematics of stochastic calculus, the
relativistic diffusion model (RDM,  \cite{Wolschin1999EPJA5,BiyajimaEtAl2002PTP108,Wolschin2016PRC94}) could thus be placed on a firm nonequilibrium-statistical foundation. 
In particular, the model has been used to analyze charged-hadron production in Pb-Pb collisions at  LHC energies, and detailed comparisons of both marginal and joint particle distributions with charged-hadron data
in transverse-momentum and pseudorapidity space from the ATLAS and ALICE collaborations have been performed. We now extend this work using an algebraic solution method that is based on a spectral eigenfunction decomposition of the Fokker--Planck operator, and promises higher accuracy at larger transverse momenta.

 {The relativistic diffusion model's scope is located between the equilibrium-statistical models for multiple hadron production that were originally proposed by Fermi \cite{fer50} and Hagedorn \cite{ha68}, and more detailed and complicated numerical nonequilibrium models that not only provide a QCD-based description of the initial stages of the collision -- such as the Color Glass Condensate (CGC, see \cite{ge10}) --, but use viscous relativistic hydrodynamics for the main part of the time evolution (e.g. \cite{koi07,luro08,alv10,hesne13,bas19,sne21}), and codes like URQMD for the hadronic final-state interactions \cite{bas98}. }

While in most of the available theories of relativistic heavy-ion collisions, charged-hadron production is considered to occur from the hot fireball of partons that expands, cools, and eventually hadronizes in a parton-hadron crossover, the RDM with three sources~\cite{Wolschin2015} also accounts for the contribution from the highly excited fragments to particle production. The role of these additional sources, which are clearly visible in net-baryon (\textit{stopping}) distributions at the Super Proton Synchrotron SPS and RHIC, has been considered in a model that relies on the interaction of valence quarks with soft gluons in the respective other nucleus and is based on gluon-saturation physics \cite{Tani2009,Tani2009b}. It has later been incorporated into a nonequilibrium-statistical approach that also accounts for the time-dependence of the stopping process \cite{Hoelck2020}. Here, the gluon-gluon source does not contribute, because it produces particles and antiparticles in equal amounts.

In particle production at energies reached at RHIC and LHC, however, the central fireball becomes the main source. The initial state is derived from quantum chromodynamics (QCD) within the color-glass model and evolved in time with a cylindrically symmetric FPE in longitudinal and transverse momentum space constructed via the appropriate fluctuation-dissipation relation (FDR), that connects the drift and diffusion coefficients with the expected asymptotic state \cite{hgw24}. The majority of the produced hadrons participate in the thermalization process, but since the system is not spatially confined, it cannot reach a true thermal equilibrium that is stationary in position space and independent of its initial state. Therefore, the asymptotic particle distribution function is expected to be a modified thermal distribution that exhibits a collective expansion of the system, reminiscent of its initial configuration. Although a correct description of the asymptotic state is crucial in the model to mesoscopically derive the form of the FPE, the system never completes the thermalization process due to a "freeze out`` of strong interactions, and charged hadrons remain far from equilibrium in the longitudinal degrees of freedom at the time they are detected. 

Analytical solutions of the relativistic diffusion model \cite{Wolschin1999EPJA5} with simplified transport coefficients have been successfully compared to LHC spectra \cite{Wolschin2015,Wolschin2016PRC94}. When solving the FPE for a two-dimensional diffusion tensor and the associated rapidity-dependent drift, however, the accuracy of the finite-difference numerical method prevented the comparison to experimental data at large transverse momenta,  where the particle yield is small compared to the peak of the distribution, but measured with high accuracy \cite{hgw24}. It is the first motivation of the present work to explore the applicability of a spectral eigenfunction solution of the FPE,
and investigate whether the method is suitable for particle production from both, fireball and fragmentation sources.  {We also investigate the effect of a dependence of the diffusion coefficient on longitudinal rapidity $y$.} The second aim is to possibly increase the numerical accuracy at larger transverse momenta through the use of this new solution method. 

 {The FPE as derived from stochastic calculus in the context of longitudinal and transverse degrees of freedom of quasi-particles in relativistic heavy-ion collisions \cite{hgw24} requires as a necessary condition for its validity the presence of a fluctuating background that is provided by the partons in the participating nucleons. Contrary to common belief, a heat bath with temperature $T$ as in classical Brownian motion of a heavy particle of mass $M$, and the corresponding restriction $M\gg T$  is not needed. As a caveat, it is mentioned that in cases where the widths of the quasi-particles become comparable to or larger than their masses -- which may be the case for pions and kaons considered in this paper --, the theoretical justification of the approach may be limited in spite of its empirical success. However, reliable calculations of the in-medium widths for light hadrons produced at LHC energies do not seem to be available.}

In the following section, the relativistic diffusion model for charged-hadron production \cite{hgw24,Wolschin1999EPJA5} in longitudinal and transverse momentum space is briefly reviewed.
We discuss the spectral method used for solving the FPE, and the applicability of the method for the central and the fragmentation sources, in Sect.\,3.
A comparison of the model results to ALICE~\cite{ALICE2013,ALICE2017,ALICE2018} and ATLAS~\cite{ATLAS2015} measurements of central Pb-Pb collisions with $\sqrt{s_\text{NN}} = 2.76$ and 5.02 TeV is given in Sect.\,4, where the free parameters are determined in $\chi^2$ minimizations. Extrapolating the parameters to $\sqrt{s_\text{NN}} = 5.36$ TeV allows us to make a prediction for the LHC Run 3 at this energy.  We discuss the results, suggest improvements of the model, and possible future outcomes of the research in the concluding Sect.\,5.
\section{Relativistic diffusion model}
Generalizing the concept of relativistic Brownian motion \cite{DebbaschEtAl1997JSP88,DunkelHaenggi2005PRE71,DunkelHaenggi2005PRE72},  we model the charged-particle trajectories in a relativistic heavy-ion collision as stochastic drift--diffusion processes \cite{hgw24}. As a consequence of the fundamental mathematical principle that stochastic processes in position space cannot be both Lorentz-invariant and (first-order) Markovian \cite{Lopuszanski1953APP12,Dudley1966AM6,Hakim1968JMP9}, the trajectories
 are expressed with respect to the Brownian particle's phase-space configuration~\(\tupdelims{\tup{x},\tup{p}}\) instead of merely its position~\(\tup{x}\). 
When including the particle's momenta~\(\tup{p}\) in the definition of the trajectory, the resulting drift--diffusion process is second-order Markovian in position-space coordinates, thus keeping the beneficial mathematical properties of Markov processes.

Denoting the probabilistic trajectories of the Brownian particle in position and momentum space as \(\tup{X} = \tupdelims{\tup{X}[1],\tup{X}[2],\tup{X}[3]}\) and \(\tup{P} = \tupdelims{\tup{P}[1],\tup{P}[2],\tup{P}[3]}\), respectively, the
increments~\(\diff{\func{\tup{X}[\mu]}{t}} \coloneq \func{\tup{X}[\mu]}{t + \diff{t}} - \func{\tup{X}[\mu]}{t}\) of the position-space trajectory due to an infinitesimal timestep~\(\diff{t}\) in the laboratory frame obey the stochastic differential equation (SDE)
\begin{equation}
	\label{eq:SDE-position}
	\diff{\func{\tup{X}[\mu]}{t}} = \frac{\func{\tup{P}[\mu]}{t}}{\func{\tup{P}[0]}{t}} \x \diff{t}
	\quad\text{for}\quad
	\mu = 0,\dots,3\,.
\end{equation}
The particle's energy is determined by the mass-shell condition
\begin{equation}
	\label{eq:SDE-mass-shell}
	\func{\tup{P}[0]}{t} \coloneq \sqrt{\textstyle m^2 + \sum_{i=1}^3 \func{\tup{P}[i]}{t}^2}
	\mathinner{,}
\end{equation}
where the \(3\)-axis is parallel to the beam axis so that \(\tup{x}[1]\) and \(\tup{x}[2]\) span the transverse plane.

The time-evolution of the particle's momentum \(\tup{P}\) is taken to follow a drift--diffusion SDE,
\begin{align}
	\label{eq:SDE-momentum}
	\diff{\func{\tup{P}[i]}{t}} =
	\func[\big]{\drift[i]{\tup{P}}}{\func{\tup{P}}{t}} \x \diff{t}
	+
	\sum_{k=1}^{3} \func[\big]{\diffusion[i k]{\tup{P}}}{\func{\tup{P}}{t}} \x \diff{\func{\tup{W}[k]}{t}}\nonumber\\
	\quad\text{for}\quad
	i = 1,2,3
	\mathinner{,}
\end{align}
with a 3-dimensional drift coefficient~\(\drift{\tup{P}}\) and a (\(3 \times 3\))-dimensional diffusion coefficient~\(\diffusion{\tup{P}}\) representing generalized directed (deterministic) and undirected (stochastic) forces, respectively.
A three-dimensional standard Wiener process~\(\func{\tup{W}}{t}\) -- the mathematical formulation of Gaussian white noise -- represents the stochastic forces.

In Eq.\,(\ref{eq:SDE-momentum}), drift and diffusion are assumed 
to be independent of the particle's position~\(\func{\tup{X}}{t}\).
The time-evolution of the system can then be accounted for as a momentum-space process~\(\tup{P}\), so that the position-space process~\(\tup{X}\) can be disregarded. 

In the present work, we only consider central heavy-ion collisions with small impact parameter, equivalent to centrality < 5\%. Here,  the system has an approximate rotational symmetry with respect to the beam axis: it is effectively two-dimensional when expressed in terms of transverse momentum~\(p_\perp\coloneq \sqrt{\parens{\tup{p}[1]}^2 + \parens{\tup{p}[2]}^2}\) and longitudinal momentum~\(p_\parallel \coloneq \tup{p}[3]\).
The transverse-plane angular coordinate can then be eliminated, resulting in a two-dimensional drift--diffusion SDE.
A hyperbolic transformation yields the coordinates transverse and longitudinal rapidity, respectively, with their connection to the transverse and longitudinal momentum
\begin{equation}
	\label{eq:rapidity-coordinates}
	h \coloneq \func[\big]{\arsinh}{p_\perp / m}
	\qquad \text{and} \qquad
	y \coloneq \func[\big]{\artanh}{p_\parallel / E}
	\mathinner{.}
\end{equation}
The transverse rapidity~\(h\) maps low \(p_\perp\) to a linear and high \(p_\perp\) to a logarithmic scale, and the particle's transverse mass is \(m \x \func{\cosh}{h} = \sqrt{m^2 + p_\perp^2} \eqcolon m_\perp\). The dynamics of the system is then determined by the
 functional form of the drift and diffusion coefficients. 
The drift coefficient can be derived from the time-asymptotic equilibrium distribution of the system.
This is conveniently done in the distribution-function representation. The probability density function (PDF) in phase space is obtained through a Kramers--Moyal expansion that yields a Fokker--Planck equation (FPE), which can be converted to a FPE in rapidity space $\xi=(h,y)$ \cite{hgw24}
\begin{equation}
	\label{eq:FPE-rapidity}
	\bracks[\big]{\derivop{t} - \func{\fpo{\Rap}}{\rap}} \x \func{\pdf{\func{\Rap}{t}}}{\rap} = 0
\end{equation}
where $\Xi=(H,Y)$ and the rapidity-space Fokker--Planck operator (FPO) is
\begin{equation}
	\label{eq:FPO-rapidity}
	\func{\fpo{\Rap}}{\rap} = \sum_{i=1}^{2} \derivop{\rap[i]} \bracks[\bigg]{- \func{\drift[i]{\Rap}}{\rap} +  \sum_{k=1}^{2} \derivop{\rap[k]} \diffusivity[i k]{\Rap} \x}
	\mathinner{.}
\end{equation}
Here, the two diffusion coefficients have been combined  into the (\(2 \times 2\))-dimensional diffusivity of \(\Rap\), which is given by
\begin{equation}
	\diffusivity[i k]{\Rap} \coloneq \tfrac{1}{2} \x \sum_{l=1}^{2} \diffusion[i l]{\Rap} \x \diffusion[k l]{\Rap}
	\quad\text{for}\quad
	i,k = 1,2
	\mathinner{.}
\end{equation}
By assuming that the system is in detailed balance, the particle distribution will approach an equilibrium state -- that need not necessarily be thermal, but can be any time-asymptotic stationary state,  \(\lim_{t \to \infty} \func{\pdf{\func{\Rap}{t}}}{\rap} \equiv \func{\pdf{\Rap_\equ}}{\rap}\). Hence, drift and diffusivity fulfill the fluctuation--dissipation relation
\begin{equation}
	\label{eq:FDR}
		\func{\drift[i]{\Rap}}{\rap} =  \sum_{k=1}^{2} \biggl[\derivop{\rap[k]} \diffusivity[ik]{\Rap} - \diffusivity[ik]{\Rap} \derivop{\rap[k]}\func{\potential{\Rap}}{\rap} \biggr],
\end{equation}
with the generalized potential
\begin{equation}
	\func{\potential{\Rap}}{\rap} \coloneq - \func[\big]{\ln}{\func{\pdf{\Rap_\equ}}{\rap}}
	\mathinner{.}
\end{equation}
If the diffusion tensor is chosen to be diagonal and constant in the rapidity variables $h$ and $y$ as in our previous work \cite{hgw24}, the FPO decomposes into a transverse and longitudinal part 
\begin{equation}
	\func{\fpo{\Rap}}{\rap}
	= D_\perp\x \derivop{h} \bracks*{\deriv{\func{\potential{\Rap}}{\rap}}{h} + \derivop{h}}
	+ D_\parallel \x \derivop{y} \bracks*{\deriv{\func{\potential{\Rap}}{\rap}}{y} + \derivop{y}}
	\mathinner{.}
\end{equation}
In the present work, we first take the rapidity-space diffusion tensor to be constant and diagonal, but shall later also consider a rapidity 
dependence of the diffusion coefficient $D_\parallel$ in longitudinal space.

It would be disadvantageous to formulate the model from the outset in $(p_\perp,y)$-space: an assumption of a diagonal diffusion tensor in this space of unlike variables
is rather unconvincing. Of course,
since collision data are usually represented as functions of transverse momenta $p_\perp$ and rapidity $y$, or pseudorapidity $\eta$, solutions of our FPE will later be converted to $(p_\perp,y)$- or $(p_\perp,\eta)$-space, respectively.

In the rapidity-space FPE, the time parameter $t$ and the diffusivity are the only dimensionful quantities. By introducing a dimensionless evolution parameter $\delta$ via $t=:t_0+\Delta t\delta$, the equation can be rewritten as 
\begin{equation}
\label{numfpe}
    \frac{\partial}{\partial \delta} f_{\bm{\Xi}}(t_0+\Delta t \delta)=\Delta t \operatorname{L}_{\bm{\Xi}} f_{\bm{\Xi}}(t_0+\Delta t \delta), \quad \delta>0,
\end{equation}
with the interaction timespan $\Delta t:=t^*-t_0$. Here, $t_0$ represents the initial time, and solving the equation up to $\delta=1$ corresponds to evolving it until the final time $t^*$.
Due to the shape of the FPO, the diffusivity coefficients and interaction timespan can be combined to form the dimensionless products $D_\perp\Delta t$ and $D_\parallel\Delta t$, whose numerical values determine how close the final state is to the initial $\left(D_\perp \Delta t, D_\parallel \Delta t \ll 1\right)$ or asymptotic state $\left(D_\perp \Delta t, D_\parallel \Delta t \gg 1\right)$. This corresponds to an intrinsic symmetry of diffusive systems: changing the interaction timespan while reciprocally adjusting drift and diffusivity leaves the final state unchanged. Mathematically, this is related to a self-similarity of the standard Wiener process $W(t)$: for any constant $\alpha>0$, the process $\widetilde{W}(t):=\frac{1}{\sqrt{\alpha}} W(\alpha t)$ is also a standard Wiener process. We shall later use Eq.\,(\ref{numfpe}) for the numerical solution of the problem.

The connection of the probability density functions of a stochastic model to the distribution of different particle species in phase space is made by scaling with the total particle number $N$. 
In relativistic heavy-ion collisions, energy and particle-number densities are highly inhomogeneous in position space, so it can be expected that particle dynamics depends on spatial coordinates as well -- whereas the subprocess in momentum space used here assumes
that its coefficient functions do not depend on positional coordinates. To address this issue, in previous applications of the model, the system has been split into multiple disconnected subsystems (\textit{sources})~\cite{Wolschin2006,Wolschin2013JPG40}
\begin{equation}
\label{eq:ndf_pdf_spurces}
   \frac{\mathrm{d}^nN}{\mathrm{d}^n q}(\bm{q}) = \sum_a N_a f_{\bm{Q}_a}(\bm{q},t^*), \quad \text{with} \quad \sum_a N_a = N,
\end{equation}
for which different initial and asymptotic states can be chosen, thereby decoupling their time evolutions with independent FPOs. Conceptually, each source occupies a distinct region in phase space, with overlap between the different regions deemed negligible. Although this approach has been proven successful \cite{Wolschin2013JPG40,hgw24}, its implementation has revealed itself particularly challenging for the solution method based on the spectral eigenfunction decomposition of the FPO that we use in the present work. The reasons will be discussed in the next chapter, along with the nature and relative contribution of each source to particle thermalization at LHC energies. In the actual calculations, only a single source will be considered, with meaningful results obtained nonetheless. 
\subsection{Initial state}
\begin{figure}
	\begin{center}
		\centering
		\includegraphics[width=1\linewidth]{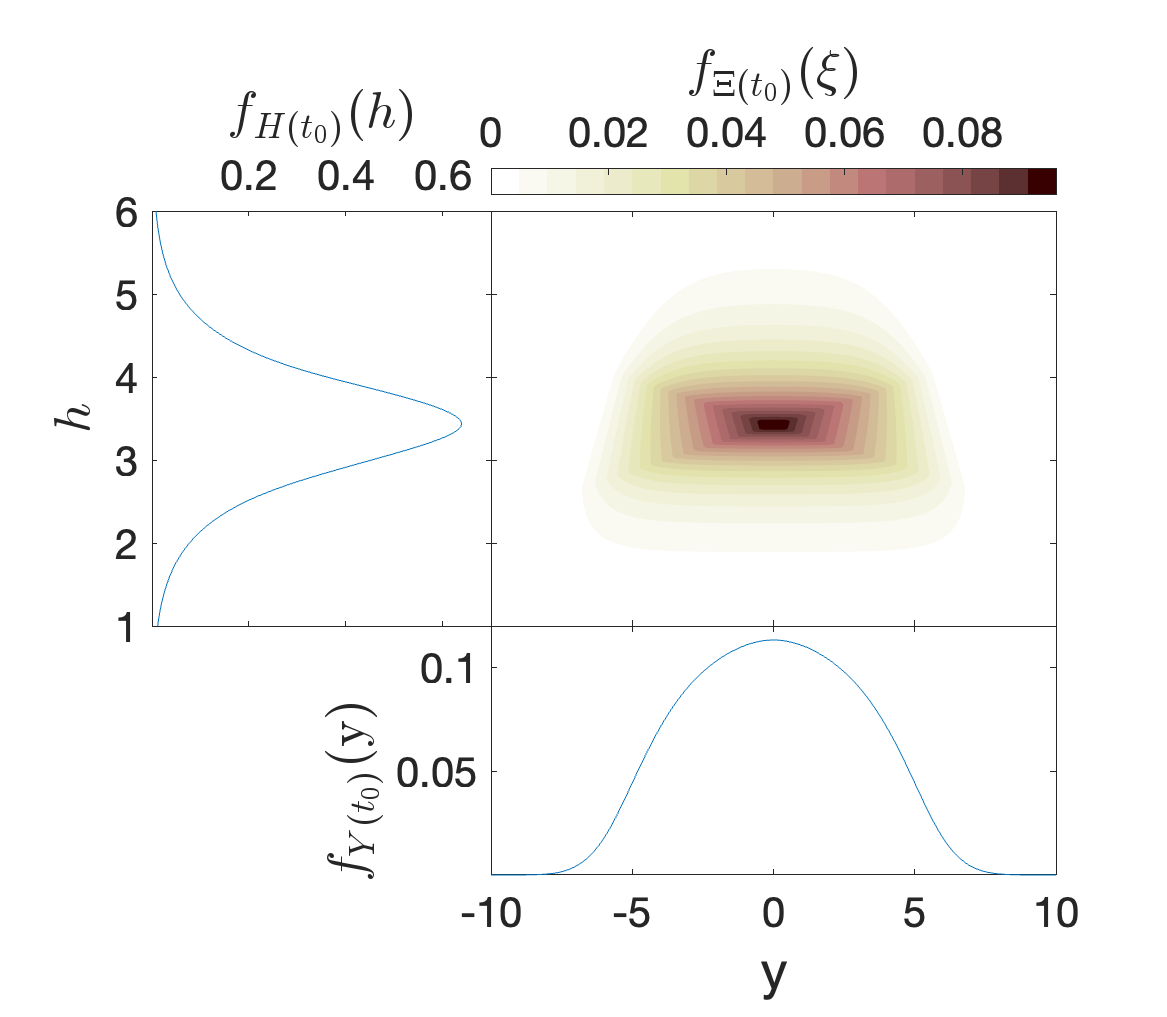}
		\caption[Initial probability density function (PDF) of the central gluon-gluon source.]{Initial probability density function (PDF) of the central gluon-gluon pion source, Eq.~\eqref{eq:pdf_initial_gg}, for a ${\sqrt{s_{\text{NN}}}= 2.76}$ TeV Pb-Pb collision. The joint distribution is shown in the central panel as a contour plot, with the color scale on the top. The left and the bottom panels contain the marginal PDFs in transverse and longitudinal rapidity, respectively.    }
		\label{fig1}
	\end{center}
\end{figure} 

The initial particle distribution functions are derived from a QCD-inspired phenomenological framework based
on gluon saturation in the {deep-inelastic scattering} of participant partons~\cite{GRIBOV1983,MUELLER1986,BLAIZOT1987,McLerran1994}, which was already successfully used \cite{Tani2009,Tani2009b,Hoelck2020} in the context of baryon stopping, and which appears to be compatible with a recent analysis of initial-state signals in ALICE data~\cite{Acharya2022}. 

Here, the key assumption is that the gluon density saturates below a characteristic momentum scale $Q_\text{s}$, so that the respective gluons form a {color-glass condensate} (CGC). For small values of the gluon Bjorken momentum fraction $x$, this scale can be parametrized as~\cite{Golec1998}
\begin{equation}
    Q_\text{s}^2(x) = \sqrt[3]{A}Q_0^2 \left( \frac{x_0}{x}   \right)^\lambda,
\end{equation}
where the values $\lambda = 0.2$ and $\frac{4}{9}Q_0^2 x_0^\lambda = 0.09$ GeV$^2$ are used to be consistent with Refs.~\cite{Hoelck2020,hgw24}. These values can be compared with literature results, where $\lambda \approx 0.288$ and $Q_0^2x_0^\lambda \approx 0.097$ GeV$^2$ were determined in a fit to deep-inelastic scattering e–p data from HERA~\cite{Golec1998}.

The interaction of the participating nucleons is assumed to be mediated by quark-antiquark pairs (dipoles) emitted by the confined partons, which then inelastically scatter off the partons of the oppositely moving nucleus. Subsequent recombination of the quarks and gluons results in the production of new hadrons, which are emitted from the nuclear fragments. From the mass $m$ and rapidities $(h,y)$ of the produced hadron, the Bjorken momentum fractions $x_\pm$ of the two generating partons contained in the forward $(+)$ and backward $(-)$ moving nucleon can be reconstructed from kinetic considerations to be
\begin{equation}
\label{eq:x_pm}
    x_\pm \approx \frac{m \cos(h) \exp(\pm y)}{2 m_\text{N} \sinh(y_\text{b})}\,,
\end{equation}
with $y_\text{b}$ the beam rapidity and $m_\text{N}$ the nucleon mass. Eq.\,\eqref{eq:x_pm} is an approximation obtained by simply imposing energy and longitudinal momentum conservation, ignoring the small initial transverse motion of the nucleons in the center-of-momentum frame, as well as the comparatively small parton masses contribution to the energy and any additional valence quarks that may be needed.

At low Bjorken $x$, the nucleon momentum is mostly carried by gluons, while valence quarks prevail at high $x$. For two interacting partons, this results in four dominant interaction processes that constitute the disconnected sources present in Eq.\,\eqref{eq:ndf_pdf_spurces}: gluon-gluon (gg), gluon-valence quark (gq), valence quark-gluon (qg) and quark-quark (qq) interactions. In this work, we consider only gluon-gluon scattering which predominantly yields hadronic particle-antiparticle pairs with low to intermediate $y$. Their initial PDF can be approximated by~\cite{KharzeevEtAl2005NPA747}
\begin{align}
    \label{eq:pdf_initial_gg}
    f_{\bm{\Xi}(t_0)}(\bm{\xi}) \approx C_{\mathrm{gg}} \sinh (h) \cosh (h)\nonumber   \qquad\qquad\qquad\\
 \times \prod_{i \in\{+,-\}} x_i \frac{G\left(x_i ; m \sinh (h)\right)}{\sinh (h)^2} \Theta(1-x_i),
\end{align}
where $G$ is the simplified gluon structure function
\begin{align}
    x G(x ; Q) \propto(1-x)^4 \min \left(Q^2, Q_{\text{s}}^2(x)\right),
\end{align}
and $C_{\mathrm{gg}}$ is a numerically determined normalization constant. Eq.~\eqref{eq:pdf_initial_gg} is completely symmetric under the exchange of the forward- and backward-going gluon, and depends on the rapidity $y$ through $x_\pm(y,h)$. The Heaviside theta functions are needed to ensure the validity of the Bjorken momentum fraction of the gluons since Eq.\,\eqref{eq:x_pm} is physically valid only for $x \leq 1$. An illustration of Eq.\,\eqref{eq:pdf_initial_gg} for a Pb-Pb collision at $\sqrt{s_{\text{NN}}}= 2.76$ TeV is shown in Fig.~\eqref{fig1}.

The hadron yield of the qq process is expected to be small compared to the other three and can thus be safely neglected, but one should not in principle ignore the contribution given by the interactions of gluons and valence quarks. These two \textit{fragmentation} sources produce particles at higher rapidities than the central source, and while essential to describe baryon stopping \cite{Hoelck2020}, their relevance in particle thermalization for central collisions at LHC energies is found to be small \cite{hgw24}. In the present work, these sources are not included because of the difficulty to account for their time evolution with the employed spectral eigenfunction decomposition, as we shall illustrate in detail when considering later the convergence properties of our algebraic solution method.

\subsection{Asymptotic state}
\begin{figure}
	\begin{center}
		\centering
		\includegraphics[width=0.9\linewidth]{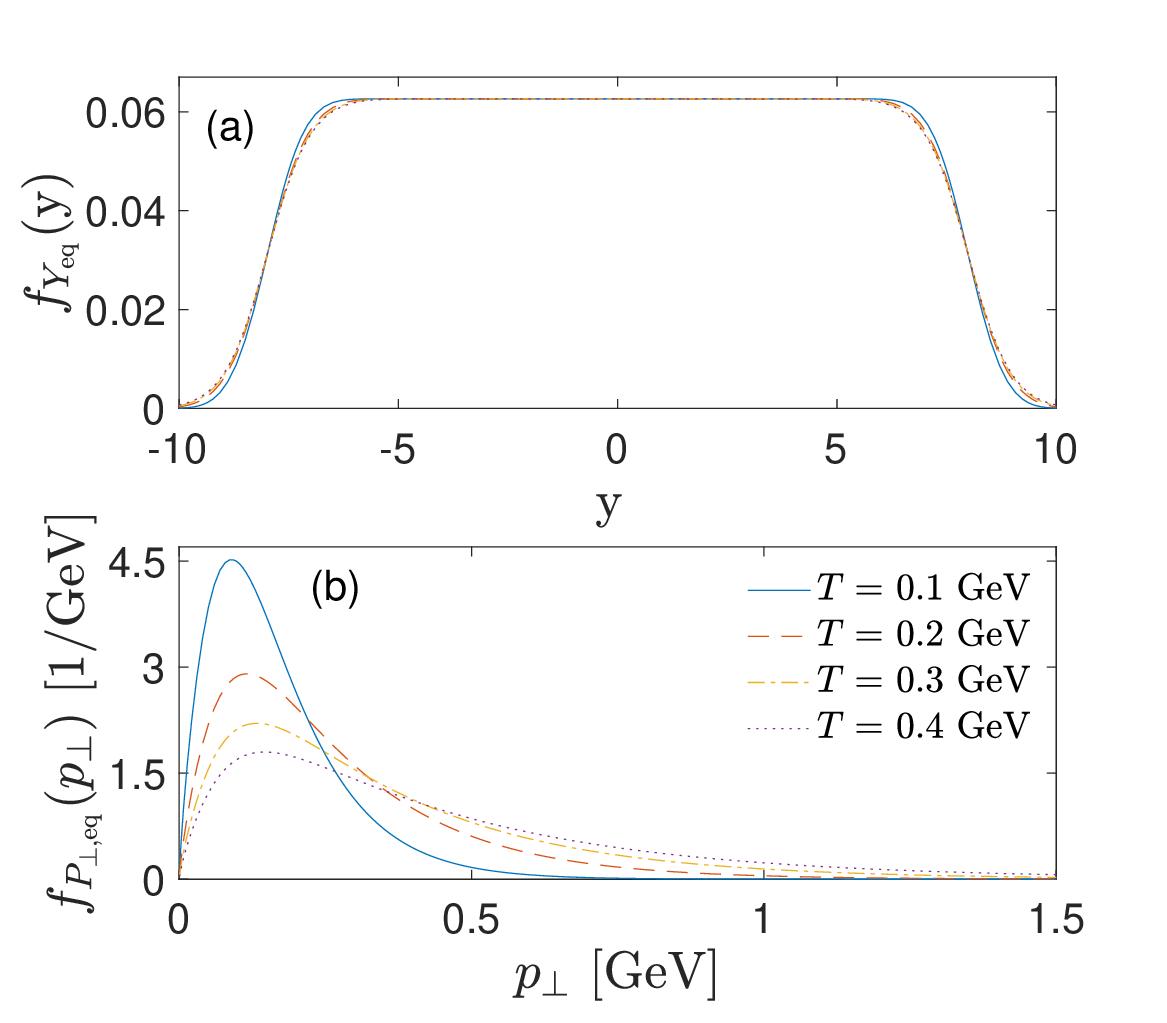}
		\caption[Asymptotic marginal PDFs for pions production.]{Asymptotic marginal PDFs for pion production in a central ${\sqrt{s_\text{NN}} = 2.76}$ TeV Pb-Pb collision for different temperatures $T$ as function of (a) $y$ and (b) $p_\perp$. Other parameters as in Table~\ref{tab1}. Considering another particle species has the same effect as changing $T$ for $f_{Y_\text{eq}}$, while greater masses move the peak of $f_{P_{\text{T},\text{eq}}}$ at higher $p_\perp$.}
		\label{fig2}
	\end{center}
\end{figure} 

If the particles produced in a relativistic heavy-ion collision maintained their interactions for an extended period without ceasing due to the physical freeze-out, it is expected that the system would undergo thermalization and reach a stationary state in momentum space. In contrast, in position space, due to the absence of spatial confinement, the system cannot reach a stationary state that is independent of its initial configuration, and it would continue to expand into the surrounding vacuum leading to a finite collective particle flow~\cite{Dunkel2007}. Consequently, the equilibrium state in momentum space of any produced charged hadron can be modeled by a generalized Maxwell-J\"uttner distribution for an expanding thermal reservoir
\begin{equation}
\label{eq:gen_therm_distr}
    f_{\bm{P}_{\text{th}}}(\bm{p}):=\frac{C_{\text{th}}}{V} \int_{\Sigma} \frac{\mathrm{d} \sigma_\mu p^\mu}{p^0} \exp \left(\frac{m-u_\nu p^\nu}{T}\right),
\end{equation}
where $\Sigma \subseteq \mathbb{R}^{1,3}$ denotes the (3+1)-dimensional space-time hypersurface containing the reservoir and $u_\nu$ its proper collective-flow velocity, which may depend on the position $\sigma \in\Sigma$. The total volume of the expanding reservoir is defined as $V := \int_{\Sigma} \mathrm{d} \sigma_\mu u^\mu $, and $C_{\text{th}}$ is a normalization constant independent of $\Sigma$ and $u_\nu$ that can be computed analytically to be
\begin{equation}
     C_{\text{th}} = \frac{\kappa}{4 \pi m^3 \exp(\kappa) K_2(\kappa)}\;,
\end{equation}
in terms of the dimensionless ratio between mass and temperature $\kappa := m/T$ and the modified Bessel function $K_a$ of the second type and order $a$.

The simplest concrete implementation of Eq.\,\eqref{eq:gen_therm_distr} is obtained by considering a non-expanding reservoir at rest,  {$u^\nu = (1,0,0,0)$}, that leads to the original Maxwell-J\"uttner distribution
\begin{subequations}
\begin{equation}
\label{eq:maxw_jutt}
     f_{\bm{P}_{\text{MJ}}}(\bm{p}) = C_{\text{th}} \exp \left(\frac{m-p^0}{T}\right),
\end{equation}
which is the relativistic generalization of the Maxwell-Boltzmann distribution proposed by J\"uttner \cite{Juttner1911} in 1911. It is useful to look at the marginal PDFs of Eq.\,\eqref{eq:maxw_jutt} in transverse and longitudinal rapidity space:
\begin{gather}
\label{eq:maxw_jutt_h}
    f_{H_{\text{MJ}}}(h) = \frac{\kappa}{K_2(\kappa)}\sinh(h)\cosh(h)^2 K_1(\kappa \cosh(h)), \\
\label{eq:maxw_jutt_y}
    f_{Y_{\text{MJ}}}(y) = \frac{1}{2 K_2(\kappa)} 
 \left[1+ \frac{2}{\kappa\cosh(y)} + \frac{2}{(\kappa\cosh(y))^2}  \right]\nonumber\\
\qquad\qquad\qquad\qquad\qquad\qquad\times \exp(-\kappa \cosh(y)).
\end{gather}
\end{subequations}
The full width at half maximum (FWHM) of the marginal rapidity distribution $\Delta Y_{\text{MJ}}$ is bounded from above by its value at the infinite-temperature limit of $2 \arcosh(\sqrt{2})$ for every finite $T$. This fact, which holds for a purely spherical flow as well, is in contrast with experimentally observed charged-hadron distributions at LHC energies that have a broader spectrum.

This problem has been addressed by considering a cylindrical-shaped expansion that spreads faster in the longitudinal than in the transverse direction: a Bjorken flow~\cite{Bjorken1983} in the longitudinal direction with maximum flow rapidity $y_s$~\cite{Schnedermann1993}
\begin{equation}
\label{eq:bjorken_flow}
    f_{\bm{\Xi}_\text{Bj}}(\bm{\xi}) = \frac{1}{2 y_s} \int_{-y_s}^{y_s} f_{\bm{\Xi}_\perp}(h,y-\zeta) \mathrm{d}\zeta
\end{equation}
on top of a transverse expansion $f_{\bm{\Xi}_\perp}$. To account for the well-known fact that conventional thermal models fail to describe the transverse high-momentum tails of charged-hadron data~\cite{Michael_1977,Hagedorn:1983}, following a phenomenological model proposed by Hagedorn \cite{Hagedorn:1983}, we have replaced \cite{hgw24} the exponential in Eq.\,\eqref{eq:gen_therm_distr} by  
\begin{equation}
\label{eq:hagedron_exp}
    \widetilde{\exp}(x;n) := \left( 1+ \frac{x}{n} \right)^n, \quad \text{with} \quad x = \frac{m-u_\nu p^\nu}{T}\,.
\end{equation}
For small values of its argument, that is for $|x/n| \ll 1$, Eq.\,\eqref{eq:hagedron_exp} behaves as an exponential function: $\widetilde{\exp} \sim \exp(x)$, whereas for $|x/n| \gg 1$, it scales as a power law: $\widetilde{\exp} \sim (x/n)^n$. This behavior captures the observed transition from a thermal distribution at low transverse momentum to a power-law decay (Pareto law) in the $p_\perp$ tail of charged-hadron spectra, which is often attributed to the increasing importance of hard processes in this regime that are not covered by thermal physics. However, when comparing to experimental data with $p_\perp > 2$ GeV, it turns out that the value of $n = -8$, adopted in our Ref.\,\cite{hgw24} for $p_\perp \le 2$ GeV following an analysis of ALICE p-p data~\cite{aamodt_production_2011}, overestimates the particle yield at high-$p_\perp$. Different values of $n$ improve the situation in the momentum tail, but only at the expense of a decreased accuracy in describing the low-$p_\perp$ region.

The agreement with high-$p_\perp$ data can be improved by considering for the asymptotic distribution an approach by Bonasera et al. \cite{Bonasera2020}, who have employed a {generalized Fokker-Plank solution} (GFPS)  to describe charged-particle transverse distributions at various LHC energies and centralities according to
 {
\begin{align}
    \label{eq:GFPS}
    f_{E_{\text{T}\text{GFPS}}}(E_\perp) = C_{\text{GFPS}}  \left[ 1+ \left(\frac{E_\perp}{b}\right)^d \right]^{-c}\nonumber\\
  \qquad\qquad\qquad\qquad\qquad \times \exp\left(-\frac{b}{\hat{T}}\arctan\frac{E_\perp}{b}\right)\,,
\end{align}
where $E_\perp := \sqrt{p_\perp^2 + m^2}-m$; $c,d$ are parameters,
 { and $\hat{T}$ is a temperature-like quantity that may differ from the equilibrium temperature $T$ because Eq.\,(\ref{eq:GFPS}) is not a thermal distribution.}  When $p_\perp \ll 1$ or $\frac{E_\perp}{\hat{T}} \ll 1$, $f_{\text{GFPS}}$ behaves as $e^{-E_\perp/\hat{T}}$,} while in the high-momentum limit it follows the power law $p_\perp^{-cd}$. Motivated by the success of Eq.\,\eqref{eq:GFPS} in describing the transverse spectra across multiple orders of magnitude, we model the asymptotic distribution of the central gluon-gluon source as 
\begin{equation}
    f_{\bm{\Xi}_\text{eq}}(\bm{\xi}) =f_{H_\text{GFPS}}(h)\; \frac{1}{2 y_s} \int_{-y_s}^{y_s} f_{{Y}_\text{MJ}}(y-\zeta) \mathrm{d}\zeta\,,
\label{eq23}
\end{equation}
which is the product of Eq.\,\eqref{eq:GFPS} expressed in terms of the transverse rapidity $h$, including the determinant of the transformation, and a Bjorken flow in $y$ based on the marginal Maxwell-J\"uttner distribution Eq.\,\eqref{eq:maxw_jutt_y}. The marginal distributions of Eq.\,(\ref{eq23}) in $y$ and $p_\perp$ for pion production are plotted in Fig.\,\ref{fig2}.

When considering the marginal rapidity distribution, $f_{{Y}_\text{eq}}(y)$ is only slightly different from $f_{{Y}_\text{Bj}}(y)$ obtained by marginalizing Eq.\,\eqref{eq:bjorken_flow}, but it improves the modeling of the transition from exponential to power-law decay in the transverse direction. In addition, the product form $f_{\bm{\Xi}_\text{eq}}= f_{H_\text{eq}} f_{Y_\text{eq}}$ greately speeds up the numerical solution. While in general such a form may not describe precisely the expected two-dimensional asymptotic state in the whole phase-space, this approximation is good enough for low-to-intermediate rapidity, where $f_{Y_\text{eq}}$ is essentially constant, as seen in Fig.\,2\,(a). Since no experimental data are available beyond this region, it is difficult to speculate where this approach starts to fail. In the next section, we proceed to the spectral solution method of the FPE.

\section{Spectral method of solution}

Spectral methods represent a powerful and versatile class of numerical techniques employed to solve partial differential and integral equations. Unlike traditional finite difference or finite element methods, the idea at the foundations of spectral methods is expressing solutions in terms of basis functions with global support. When applied to a smooth problem on a simple domain, spectral methods can achieve exceptional accuracy compared to other numerical methods, and the convergence is usually exponential in the number of basis functions or evaluation points \cite{Shizgal2015,trefethen_spectral_2000}.

We first discuss the use of non-classical orthogonal polynomials to efficiently construct a spectral representation of the FPO. Using such a basis set, it is possible to implement a representation of the one-dimensional operator that does not require an explicit calculation of the drift coefficient via the FDR, while for the higher-dimensional case, the equivalence between the Fokker-Planck and a Schrödinger equation can be exploited.

To construct a non-classical basis set, consider the polynomials $\{P_\text{m}\}$ orthonormal with respect to some weight function $w$ on $(\mathrm{a}, \mathrm{b}) \in \mathbb{R}$, that is,
\begin{equation}
    \label{eq:ort_pol}
    \int_a^b w(x) P_\text{m}(x) P_\perp(x) \mathrm{d} x=\delta_{\text{ml}} .
\end{equation}
The polynomials form a basis set and satisfy the general three-term recurrence relation~\cite{davis2014}
\begin{equation}
    x P_\text{m}(x)=\sqrt{\beta_{\text{m}}} P_{\text{m}-1}(x)+\alpha_{\text{m}} P_\text{m}(x)+\sqrt{\beta_{\text{m}+1}} P_{\text{m}+1}(x),
\end{equation}
where the coefficients $\alpha_{\text{m}}$ and $\beta_{\text{m}}$ can be determined with the Stieltjes procedure \cite{GAUTSCHI1985}. 
\subsection{Spectral representation of the 1D FPO}

The one-dimensional FPO can be converted into a Hermitean form by multiplying it with $e^{\Phi}$, that is, by dividing it by the stationary solution $f_{X_\text{eq}}$\footnote{$f_{X_\text{eq}}$ is assumed to be normalized, such that all normalization factors are ignored in this section.}.
Writing the general time-dependent solution $f_{X}(x,t)$ as the product of the stationary solution and an auxiliary function $g_X(x,t)$
\begin{equation}
    f_{X}(x,t) := f_{X_\text{eq}}(x)g_{X}(x,t),
\end{equation}
the FPE for the function $g_X$ becomes 
\begin{equation}
\label{eq:fpe_g}
    \begin{aligned}
\frac{\partial g_X}{\partial t} & =\frac{1}{f_{X_\text{eq}}} \frac{\partial}{\partial x}\left[D_X f_{X_\text{eq}} \frac{\partial g_X}{\partial x}\right]  \\
& =\mu_X \frac{\partial g_X}{\partial x}+D_X \frac{\partial^2 g_X}{\partial x^2}  =\operatorname{G}_Xg_X,
\end{aligned}
\end{equation}
where $G_X$ is self-adjoint with respect to the equilibrium solution as a weight function
\begin{equation}
\begin{aligned}
        \langle \phi_1 |\operatorname{G}_X | \phi_2 \rangle &= \int f_{X_\text{eq}}  \phi_1  \operatorname{G}_X  \phi_2 \mathrm{d}x \\
        &=\int f_{X_\text{eq}}  \phi_2  \operatorname{G}_X  \phi_1 \mathrm{d}x\\
        &= \langle \phi_2 |\operatorname{G}_X | \phi_1 \rangle\,,
\end{aligned}
\end{equation}
provided the zero flux boundaries condition
\begin{equation}
    \int_{\partial V} f_{X_\text{eq}}  D_X  \frac{\partial g}{\partial x} \mathrm{d}x =0\,.
\end{equation}
To construct an efficient spectral representation of the operator $\operatorname{G}_X$, a basis of non-classical polynomials $\{P_\text{m}\}$ orthogonal w.r.t. the stationary solution of the FPE can be employed. Such a basis satisfies Eq.\,\eqref{eq:ort_pol} with $w = f_{X_\text{eq}}$, and has the advantage that it is automatically adapted to the specific problem. In addition, knowing that the first eigenfunction of the FPO is the equilibrium distribution and that the successive eigenfunctions are orthogonal to it, a basis set satisfying the same orthogonality relations is special compared to other polynomial sets.
The spectral representation of $\operatorname{G}_X$ in this basis is \cite{shizgal1985,shizgal1996,Shizgal1998} 
\begin{equation}
\begin{aligned}
    \operatorname{G}_X^{\text{ml}} & =\int f_{X_\text{eq}} P_\text{m} \operatorname{G}_X P_\perp \mathrm{d} x \\
    & =\int f_{X_\text{eq}} P_\text{m} \frac{1}{f_{X_\text{eq}}} \frac{d}{d x}\left[f_{X_\text{eq}} D_X \frac{d P_\perp}{d x}\right] \mathrm{d}x\,.
\end{aligned}
\end{equation}
Integrating by parts and making use of the orthogonality of the basis set w.r.t. the stationary solution, the expression simplifies to 
\begin{equation}
\label{eq:G_spec}
    \operatorname{G}_{X}^{\text{ml}}=\int f_{X_\text{eq}} D_X P_\text{m}^{\prime}P_\perp^{\prime}\mathrm{d}x,
\end{equation}
where natural boundary conditions were assumed. This representation is particularly convenient because it only requires polynomial derivatives that can be carried out analytically. Since the drift function does not explicitly appear in the expression, if one already knows the stationary solution and the form of the diffusion coefficient, the use of the FDR is not even necessary. The eigenvalues $\lambda_\text{m}$ and eigenfunctions $\psi_\text{m}$ of the operator $\operatorname{G}_{X}$ can be estimated by diagonalization of the $(M+1)\times(M+1)$ symmetric matrix $\operatorname{G}_{X}^{\text{ml}}$, where $M+1$ is the dimension of the orthonormal basis set $\{P_\text{m}\}$. 
It was shown numerically that the coefficients of the expansion of $\psi_\text{n}$ in the $\{P_\text{m}\}$ basis set are linear variational parameters. The variational theorem then guarantees that this method provides an upper bound to the eigenvalues for each $M$ and converges from above.\\
The FPO ${L}_{X}$ and $\operatorname{G}_{X}$ share the same eigenvalues, and are connected trough 
\begin{equation}
    \operatorname{G}_{X} = e^{\Phi} \operatorname{L}_{X} e^{-\Phi}.
\end{equation}
The time-dependent solution of the FPE can be then approximated by 
\begin{equation}
\label{eq:fpe_exp_N}
    f_{X}(x,t) \approx  f_{X_\text{eq}}(x) \sum_{\text{m}=0}^{M} c_\text{m} \psi_\text{m}(x) e^{-\lambda_\text{m} (t-t_0)}\,, 
\end{equation}
where the coefficients $c_\text{m}$ are determined from the initial condition by
\begin{equation}
	\label{eq:exp_coeff}
    c_\text{m} = \int f_{X}(x,t_0) \psi_\text{m}(x) \mathrm{d}x\,.
\end{equation}
The goodness of Eq.\,\eqref{eq:fpe_exp_N} in approximating $f_{X(t)}$ depends of course on the convergence of the eigenvalues and eigenfunctions of the FPO, and on the convergence of the coefficients $c_\text{m}$ to small values when approaching the $M$-th element.

\subsection{Representation of the 2D Schrödinger operator}
The stochastic process Eq.\,(\ref{eq:FPE-rapidity}) is characterized by a two-dimensional FPO, for which the simple 1D representation described above cannot be employed. 
The equivalence between the FPE and a Schr\"odinger equation with a supersymmetric potential is a well-understood fact~\cite{Risken1996,Shizgal2016}, and it is useful to take advantage of the large class of solution methods of quantum mechanical problems. Finding a spectral solution of the FPO can then be reduced to the eigenvalue equation of a 2D Hamiltonian
where the Schr\"odinger potential can be obtained from the drift and diffusion coefficient functions. An explicit transformation in the case of constant diagonal diffusion is given in Ref.~\cite{darboux}. 

Unfortunately, it is not generally possible to construct an orthonormal set of polynomials of two variables $(x,y)$ w.r.t. some weight function $w(x,y)$, and two different basis sets have to be employed to construct a spectral representation~\cite{shizgal1996}. 
In contrast to the one-dimensional representation Eq.\,\eqref{eq:G_spec}, the explicit calculation of the drift coefficient and its derivative is necessary to obtain the Schr\"odinger potential. 
If $\mu_{\bm{X}}$ is constructed from the FDR, that means carrying out second derivatives of the asymptotic PDF. As explained in Sect.\,4, it will be sufficient to solve an effective 1D FPE for our case of interest, and we can then avoid the use of this 2D representation.
\subsection{Convergence of the spectral expansion}
\label{sec:convergence}
\begin{figure}
	\begin{center}
		\centering
		\includegraphics[width=0.9\linewidth]{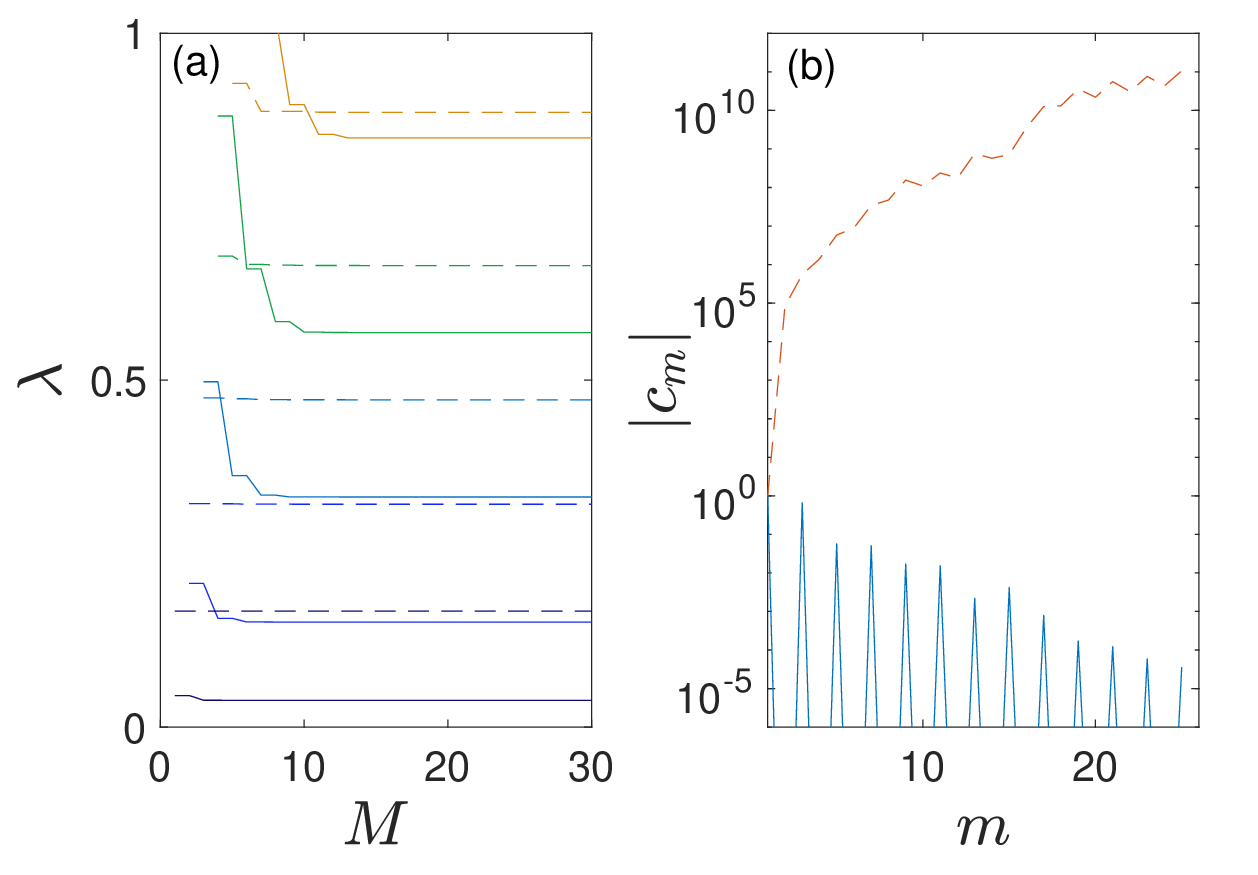}
		\caption[Spectral convergence of the FPE.]{Spectral convergence of the FPE for the central gluon-gluon (solid curves) and the fragmentation sources (dashed curves). (a) The first five positive eigenvalues of the adimensional marginal rapidity FPO are plotted against the orthogonal basis dimension $M$. The eigenvalues of the fragmentation FPO are scaled by 0.18 for better visualization. (b) Absolute value of the expansion coefficients $c_\text{m}$ (logarithmic scale) calculated from the initial PDF of the central (solid) and forward (dashed) sources. $M=25$.}
		\label{fig3}
	\end{center}
\end{figure}
\begin{figure}
	\begin{center}
		\centering
		\includegraphics[width=0.8\linewidth]{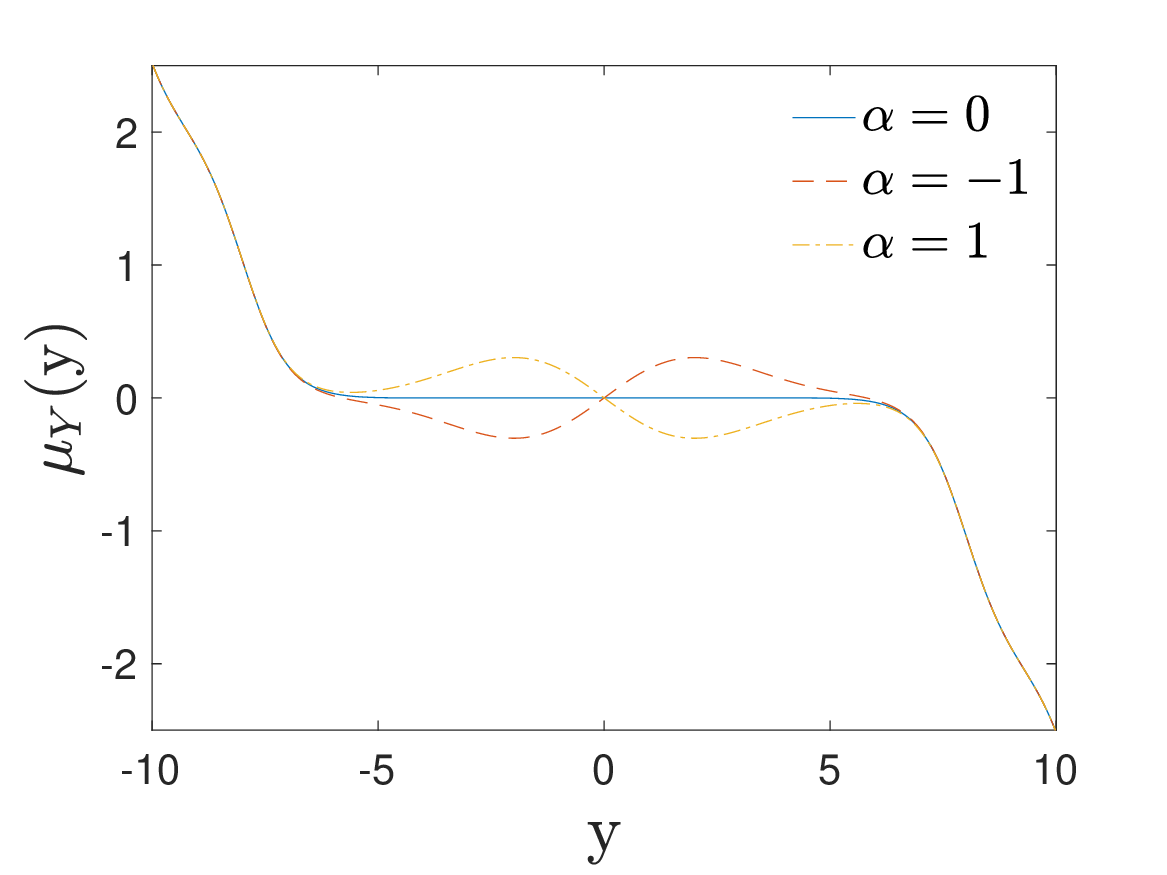}
		\caption[Drift coefficient $\mu_Y$ for three different values of $\alpha$]{Drift coefficient $\mu_Y$ for three different values of $\alpha$. The $f_{Y_\text{eq}}$ in the FDR describe pion thermalization in  ${\sqrt{s_\text{NN}} = 2.76}$ TeV Pb-Pb collisions. Parameters as in Table~\ref{tab1}, except for $D_\parallel \Delta t = 1$. The solid blue line, corresponding to $\alpha = 0$, represents the case of a constant diffusion. The dashed red line and the dot-dashed yellow line are calculated for $\alpha = -1$ and  $\alpha = 1$, respectively. }
		\label{fig4}
	\end{center}
\end{figure}
We limit the discussion of the FPE spectral solution convergence to the marginal one-dimensional rapidity FPO
\begin{equation}
	\operatorname{L}_Y = D_{\parallel} \frac{\partial}{\partial y}\left[-\frac{\partial \ln f_{Y_\text{eq}}(y)}{\partial y}+\frac{\partial}{\partial y}\right],
\end{equation}
which can be solved employing the representation Eq.\,\eqref{eq:fpe_exp_N} after the transformation to its Hermitean form $\operatorname{G}_Y$. In Fig.~\ref{fig3}, the extremely fast convergence of this solution scheme is shown when considering the FPO generated by the marginal rapidity central gluon-gluon source asymptotic PDF for hadron thermalization in a Pb-Pb collision at $\sqrt{s_\text{NN}} = 2.76$ TeV with constant diffusion $D_\parallel$. As inferred from Fig.~\ref{fig3}\,(a), using an orthogonal basis of dimension $M=15$ is already sufficient to obtain a good convergence of the first five positive eigenvalues. Given this fact, for the subsequent results presented in this work, a basis dimension of $M=25$ is employed, which is deemed sufficient for obtaining a satisfactory precision of the relevant FPE solution. Panel (b) shows on a logarithmic vertical scale the exponential approach to zero of the expansion coefficients $c_\text{m}$, calculated using Eq.\,\eqref{eq:exp_coeff} with the initial PDF of the central gg source Eq.\,\eqref{eq:pdf_initial_gg} and the eigenfunctions of $\operatorname{G}_Y$ calculated with $M=25$. The values of the coefficients corresponding to odd eigenfunctions are zero up to machine precision, as expected by the $y$-parity of both the FPO and the initial condition. 

In contrast to the success of this spectral solution scheme in solving the central-source FPE, Fig.~\ref{fig3} also illustrates its failure when applied to the forward quark-gluon and backward gluon-quark fragmentation sources. In panel (a), the dashed lines show that the convergence of the eigenvalue is even faster than for the gg source, but from panel (b) it can be observed that the absolute values of $c_\text{m}$ explode for the higher eigenfunctions, and therefore a good approximation of the initial state with a finite-dimensional polynomial basis orthogonal to $f_{Y_{\text{eq}}}$ cannot be reached. To understand why that is the case, the fragmentation equilibrium and initial states rapidity dependence have to be analyzed. Similarly to the central fireball, the marginal longitudinal asymptotic state is given by a $y$-symmetric Bjorken flow Eq.\,\eqref{eq:bjorken_flow} for both the gq and qg sources. The  (backward)forward source initial PDF is again derived from the DIS-CGC- formalism, and is peaked around (-)$y_\text{b}$. 
It turns out that  the forward qg marginal rapidity asymptotic and initial PDF -- as employed in Ref.\,\cite{hgw24} --
occupy two almost disconnected regions in rapidity space. At the peak of $f_Y(t_0)$, the extremely fast decay of $f_{Y_\text{eq}}$ has already reached the machine noise level. While it is theoretically true that an infinite-dimensional orthogonal basis set defined by Eq.\,\eqref{eq:ort_pol} on $\mathbb{R}$ is complete, and can then approximate any continuous function, in practice one can only generate a finite set. Consequently, the method is not suited in practice to calculate the time evolution of the fragmentation sources.


\begin{table*}
    \centering
    \begin{tabular}{c|c|c|c|c|c|c|c|c}
    \hline
   $\sqrt{s_{\text{NN}}}$\,(TeV) & $y_\text{b}$ & $N \times 10^{-4}$ & $D_\parallel\Delta t$ &  $\alpha$ & $\hat{T}\,(\text{MeV})$ & $b$\, {(GeV)}  & $cd$ & $\chi^2/\text{ndf}$\\
         \hline
         2.76&  7.99&  1.713(8)&  0.98(9)& -1&  486(4)&  9.4(5)&  0.30(2)& 134/193\\
    
         5.02&  8.58&  2.182(8)&  $1.055^*$& -1&  624(7)&  22(4)&  2.8(2)& 28/66\\
         
         5.36&  8.65&  $2.237^*$&  $1.063^*$& -1$^*$&  639$^*$&  24$^*$&  3.06$^*$& -\\
         \hline
    \end{tabular}
    \caption{Model parameters for central Pb-Pb collisions determined in $\chi^2$ minimizations to ALICE and ATLAS data, and extrapolation to $\sqrt{s_\text{NN}} =5.36$ TeV. Numbers in brackets indicate the estimated uncertainty of the preceding digit; see text for details, and for the physical meaning of  {the temperature-like parameter} $\hat{T}$. Numbers marked with $^*$ are extrapolated as explained in the text. The value of $c$ is fixed to 5. }
    \label{tab1}
\end{table*}
\section{Comparison with LHC data}

For $^{208}$Pb-Pb collisions at LHC energies, the only available particle-production data in the rapidity region up to $y=5$ are for unidentified charged hadrons $h^ \pm$, because no suitable spectrometer for hadron identification has been installed yet at $|y|>2$. Consequently, the measured distributions are given in terms of the coordinates $\left(p_{\perp}, \eta\right)$, where $\eta$ is the {pseudorapidity}
\begin{subequations}
\begin{equation}
    \eta:=\operatorname{artanh}\left(\frac{p^3}{|\bm{p}|}\right)=\operatorname{arsinh}\left(\sqrt{1+\left(\frac{m }{p_{\perp}}\right)^2} \sinh (y)\right),
\end{equation}
 {and the Jacobian matrix for the transformation from $(p_\perp,y)$ to $(p_\perp,\eta)$ space is given by}
\begin{equation}
    \frac{\partial\left(p_{\perp}, \eta\right)}{\partial\left(p_{\perp}, y\right)}=\left(\begin{array}{cc}
    1 & 0 \\
    -\frac{\tanh (\eta)}{p_{\perp}\left[1+\left(\frac{p_{\perp}}{m }\right)^2\right]} & \sqrt{1+\left(\frac{m }{p_{\perp} \cosh (\eta)}\right)^2}
\end{array}\right).
\end{equation}
\end{subequations}
The measurement of $\eta=-\ln[\tan({\theta/2})]$ requires only the knowledge of the scattering angle $\theta$, which is available without particle identification. The pseudorapidity agrees with $y$ for $|\bm{p}| \gg m $.
The charged-particle yield is mainly composed of pions, followed by kaons and (anti)protons, with the measured abundance fractions of approximately 83\%, 13\%, and 4\%, respectively, in 2.76 TeV Pb-Pb ~\cite{ALICE2013_b,ALICE2020}. 

Comparison with data from ALICE~\cite{ALICE2013,ALICE2017,ALICE2018} and ATLAS~\cite{ATLAS2015} of Pb-Pb collisions with $\sqrt{s_\text{NN}} = 2.76$ and 5.02 TeV is performed by evolving the initial PDFs of the three hadron species separately with a common value of $\bm{D}_{\bm{\Xi}} \Delta t$ in the $(h,y)$ coordinate space. Then, the obtained PDFs are transformed to the $(p_\perp,\eta)$ space with the appropriate determinant, multiplied by the total number of particles $N$, and finally summed according to the above percentages. While in principle every particle species could have an independent set of parameters, the lack of identified hadron spectra would make the then abundant parameter space difficult to fit meaningfully to the available data. For this reason, the initial and final PDFs for pions, kaons, and protons are assumed to differ only by the respective particle masses, $m_\pi =  139.6$ MeV, $m_K = 493.7$ MeV, and $m_p = 938.3$ MeV, and share the rest of the free parameters. 

The parameter space can be additionally reduced by adopting two results from our Ref.\,\cite{hgw24}. There, we have identified the maximum Bjorken flow rapidity $y_s$ for the central gluon-gluon source with the beam rapidity $y_\text{b}$, and it has been verified that the equilibration in the transverse direction happens much faster than in the longitudinal one, corresponding to $D_\perp \Delta t \gg D_\parallel \Delta t$. While this is not physically surprising, as thermal models with modified high-momentum tails are known to well describe transverse charged-hadrons spectra, it affects negatively the precision and speed of the numerical solution of the FPE. Following the same approach as in Ref.\,\cite{hgw24}, the problem is circumvented by taking the limit $D_\perp \Delta t \to \infty$, which in practice means assuming an almost instantaneous thermalization in the transverse direction. 
Hence, the problem reduces to solving only the marginal one-dimensional longitudinal FPE
\begin{equation}
	\frac{\partial f_Y}{\partial t} = \left[ -\frac{\partial}{\partial y} \mu_Y + \frac{\partial^2}{\partial y^2} D_Y \right] f_Y,
\end{equation}
which can be done with the efficient spectral representation Eq.\,\eqref{eq:G_spec}. The two-dimensional joint distribution can then be reconstructed by
\begin{equation}
\begin{aligned}
	f_{\bm{\Xi}}(\bm{\xi},t^*) &\approx f_{Y}(y,t^*) f_{H_\text{eq}|Y_\text{eq}}(h|y) \\
	& =  \frac{f_{Y}(y,t^*)}{f_{Y_\text{eq}}(y)} f_{\bm{\Xi}_\text{eq}}(\bm{\xi})\,.
\end{aligned}
\end{equation}

Since the explicit use of the FDR is not necessary in this simplified case, and being motivated by the difficulty of a constant longitudinal diffusion coefficient to reproduce the midrapidity region encountered previously \cite{hgw24}, it is straightforward to allow for a rapidity-dependent diffusion $D_Y(y)$. Imposing that $D_Y$ is continuous, differentiable, positive $\forall y$ and respects the  even symmetry of the system, a simple form is given by
\begin{equation}
    D_Y(y) = D_\perp ( 1 + \alpha e^{-y^2/y_\text{b}}), \quad \alpha > -1,
\end{equation}
where the denominator of the exponential is chosen to be $y_\text{b}$ as it is the only fixed rapidity scale of the system, and treating it as a free parameter does not improve significantly the fit to the data. Since the solution obtained with a constant diffusion predicted a particle yield around $y=0$ too low compared to the data, and because in that region the PDF is monotonically decreasing in the time evolution, it is expected that negative values of $\alpha$ should improve the modeling of hadron spectra, as will be shown by the fit to LHC data in the next subsection. The corresponding drift coefficient for positive and negative values of $\alpha$ is shown in Fig.\,\ref{fig4}.
\begin{figure*}
	\begin{center}
		\centering
		\includegraphics[width=\linewidth]{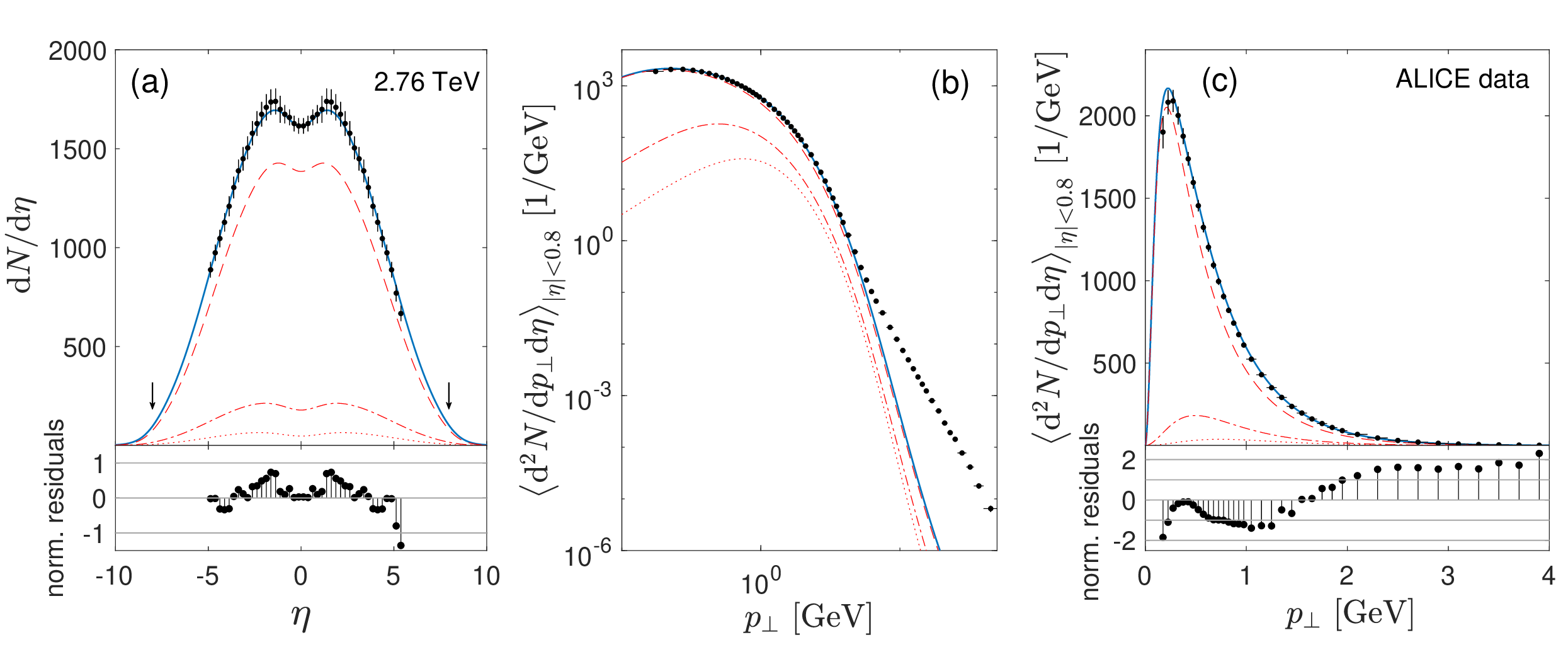} 
		\caption[Comparison of the calculated pseudorapidity spectrum to ALICE data for ${\sqrt{s_\text{NN}} = 2.76}$ TeV.]{(a) Comparison of the calculated marginal pseudorapidity charged-hadrons number density function $\mathrm{d}N/\mathrm{d}\eta$ of central $\sqrt{s_\text{NN}} = 2.76$ TeV Pb-Pb collisions to 0-5\% ALICE data~\cite{ALICE2013}. In the top panel, the solid blue curve indicates the total calculated particle yield, while the red dashed, dot-dashed, and dotted lines are the contributions of pions, kaons, and (anti)protons, respectively. The black dots are the experimental data points, with the vertical error bars including statistic and systematic error. The bins are smaller than the symbol size, and the black arrows mark the beam rapidity. The bottom panel shows the normalized residuals, as defined in the text, for every bin. (b) Comparison of the calculated joint produced charged-hadron spectrum $\mathrm{d}^2N/\mathrm{d}\eta\mathrm{d}p_\perp$ to 0-5\% ALICE data~\cite{ALICE2018} in $\sqrt{s_\text{NN}} = 2.76$ TeV Pb-Pb central collisions, averaged over the $|\eta| < 0.8$ slice and plotted against $p_\perp$ on a log-log scale. The four curves represent the same quantities as in (a). Horizontal bars indicate the bin size, and experimental error bars are smaller than the symbols. See text for the deviation beyond $p_\perp = 4$ GeV. (c) Same as in (b), but restricted to the interval $p_\perp < 4$ GeV and on a linear scale. Normalized residuals are shown in the bottom panel.}
		\label{fig5}
	\end{center}
\end{figure*}
\begin{figure}
	\begin{center}
		\centering
		\includegraphics[width=0.94\linewidth]{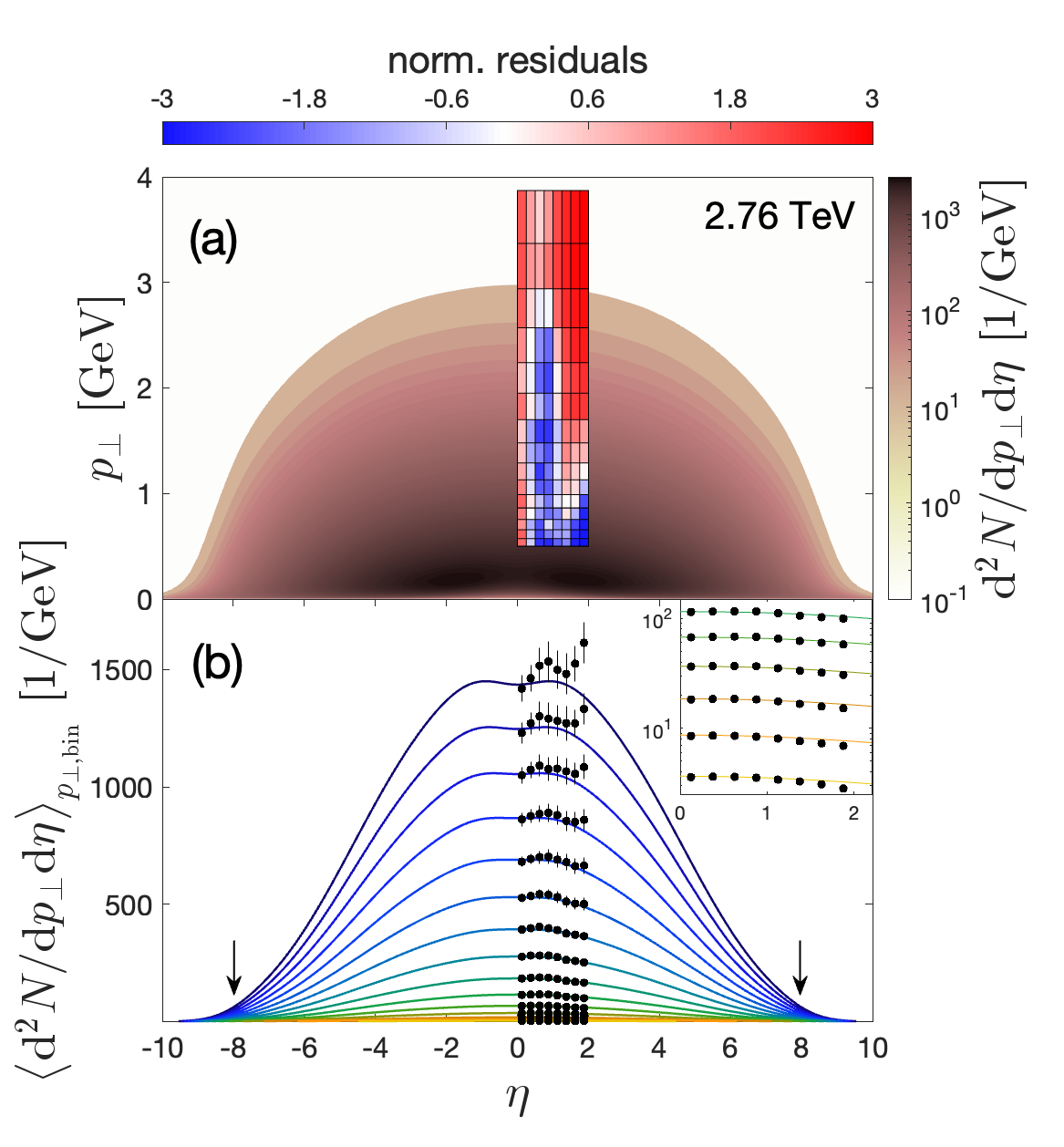} 
		\caption[Joint two-dimensional spectrum compared to ATLAS data for ${\sqrt{s_\text{NN}} = 2.76}$ TeV.]{(a) Joint two-dimensional spectrum of produced charged hadrons in ${\sqrt{s_\text{NN}} = 2.76}$ TeV Pb-Pb central collisions, compared to 0-5\% ATLAS data~\cite{ATLAS2015}. The logarithmic color scale on the right refers to the calculated value of $\mathrm{d}^2N/\mathrm{d}\eta\mathrm{d}p_\perp$, and the linear color scale at the top refers to the normalized residuals -- data minus model prediction divided by total uncertainty -- represented as rectangles the size of the corresponding experimental bin. (b) Same joint spectrum, but averaged over experimental $p_\perp$ bins from  $p_\perp =0.5$ GeV (top curve) to  $p_\perp = 3.87$ GeV (bottom curve) and plotted against positive $\eta$. The insert shows in detail the six lowest curves, from  $p_\perp =1.7$ GeV to  $p_\perp = 3.87$ GeV, on a vertical logarithmic scale. Vertical and horizontal bars indicate the total uncertainties and the bin sizes in the $\eta$ direction. The black arrows mark the beam rapidity. }
		\label{fig6}
	\end{center}
	\end{figure}


\begin{figure*}
	\begin{center}
		\centering
		\includegraphics[width=\linewidth]{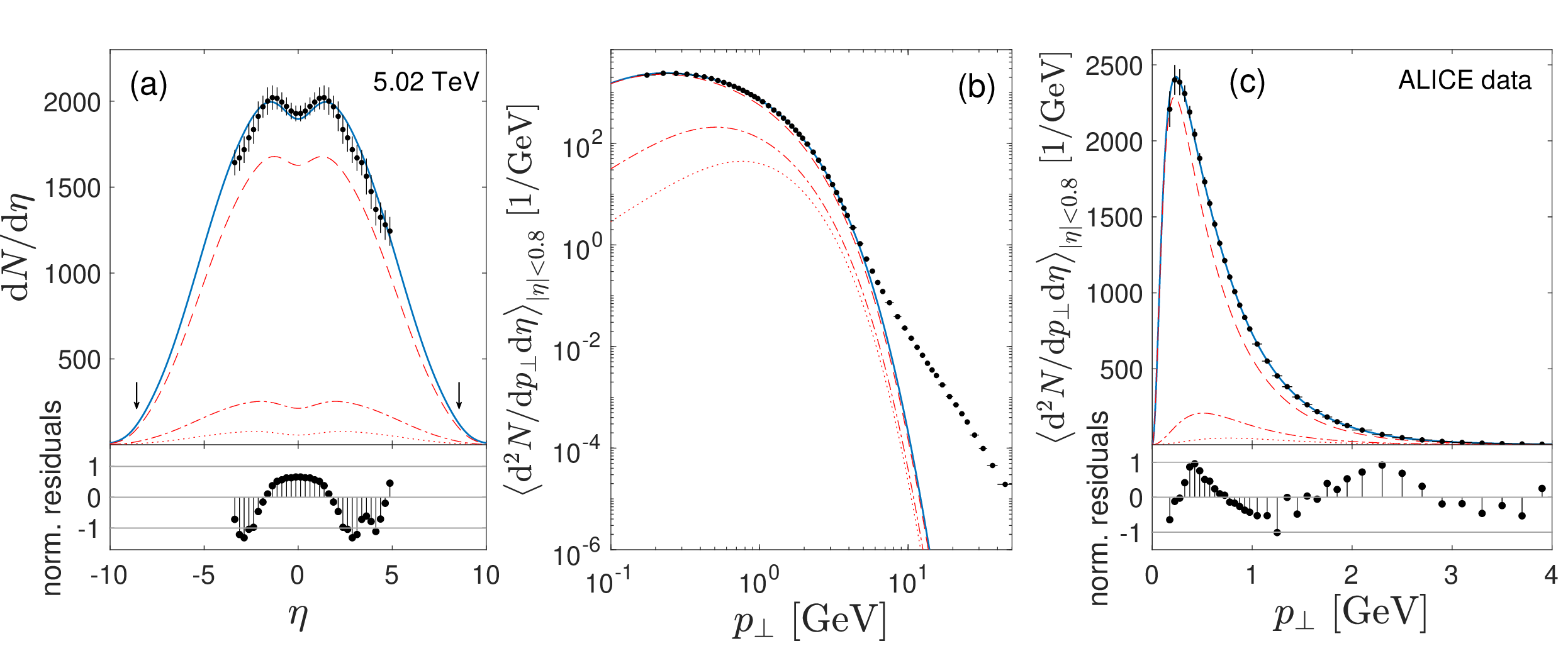} 
		\caption[Comparison of the calculated pseudorapidity spectrum to ALICE data for ${\sqrt{s_\text{NN}} = 5.02}$ TeV.]{(a) Comparison of the calculated marginal pseudorapidity charged-hadron number density distribution $\mathrm{d}N/\mathrm{d}\eta$ of central $\sqrt{s_\text{NN}} = 5.02$ TeV Pb-Pb collisions to 0-5\% ALICE data~\cite{ALICE2017} (black points). The top panel shows the total produced hadrons (solid curve) and the contribution of pions (dashed), kaons (dot-dashed), and (anti)protons (dotted). Horizontal bars represent the total experimental errors, the bins are smaller than the symbol size, and the black arrows mark the beam rapidity. The bottom panel shows the normalized residuals for every bin. (b) Comparison of the calculated joint produced charged-hadron spectrum $\mathrm{d}^2N/\mathrm{d}\eta\mathrm{d}p_\perp$ to 0-5\% ALICE data~\cite{ALICE2018} in $\sqrt{s_\text{NN}} = 5.02$ TeV Pb-Pb central collisions, averaged over the $|\eta| < 0.8$ slice and plotted against $p_\perp$ on a log-log scale. See text for deviation above 4 GeV. (c) Same as in (c), but restricted to the interval $p_\perp < 4$ GeV and on a linear scale. Normalized residuals are shown in the bottom panel.}
		\label{fig7}
	\end{center}
\end{figure*}

Allowing for a rapidity-dependent diffusion, the free parameters of the model are $D_\parallel \Delta t$ and $\alpha$ that determine how much and how the solution approaches equilibration, the total number of particles $N$, and the parameters of the equilibrium PDFs. These are $\hat{T}$, $b$, and the exponents $c$ and $d$.  
The fit to experimental data is based on a least-$\chi^2$ determination. For every data point, the calculated PDF is integrated in the respective coordinate bin, and the result is subtracted from the experimental value and divided by the total uncertainties, statistic and systematic. These normalized residuals are squared and summed over all available data at each collision energy, and the resulting total $\chi^2$ is minimized. Since the aim of the present work is to describe particle production in relativistic heavy-ion collision with a non-equilibrium statistical model, comparison with data is carried out only in the $p_\perp \leq 4$ GeV region, outside which hard partonic interactions are expected to dominate over statistical effects. At high transverse momentum, a non-equilibrium statistical description such as the one employed in this work cannot be expected to hold, and perturbative QCD approaches as in Refs.\,\cite{BAIER1997,Guo2000,Burke2014} have to be applied. The minimization process is carried out with the \textsc{Matlab} routine \texttt{fminsearch.m}, based on the Nelder-Mead simplex algorithm~\cite{Nelder_mead}. Uncertainties of the parameter vector $\bm{\theta}$ are estimated with the square root of the covariance matrix diagonal, which is obtained by inverting the Hessian of $\chi^2(\bm{\theta})$. The latter is calculated numerically with the \textsc{Matlab} suite \texttt{DERIVEST}~\cite{Derivest}.
\subsection{Central $\sqrt{s_\text{NN}} = 2.76$ TeV Pb-Pb} 
The results of the parameter optimization for both LHC energies are shown in Table~\ref{tab1}. While the transverse thermalization is almost instantaneous, the system remains very far from equilibrium in the longitudinal direction, as indicated by the relatively small values of $D_\parallel \Delta t$. 
The $\chi^2$ optimization reaches the lower bound of $\alpha = -1$ for both $\sqrt{s_\text{NN}} = 2.76$ TeV and $5.02$ TeV, corresponding to an almost vanishing equilibration at $y=0$. As already argued in Ref.\,\cite{Bonasera2020}, it was verified that the parameters describing the transverse FPE solution are highly correlated. In particular,  only the value of the product $cd$ is relevant when stopping the fit at $p_\perp = 4$ GeV, and the degrees of freedom of the model were then reduced from seven to six.  The high values of the parameter $\hat{T}$ introduced in Ref.\,\cite{Bonasera2020} are due to the fact that -- different from the temperatures determined in Ref.\,\cite{hgw24} -- it should not be identified with the physical temperature of the system, because the GFPS is not a (modified) thermal distribution.
For $\sqrt{s_\text{NN}} = 2.76$ TeV, the ALICE data set~\cite{ALICE2013} contains measurements of the marginal pseudorapidity PDF in the $-5 \leq \eta \leq 5.5$ window. The comparison of our model results with the data is presented in Fig.~\ref{fig5}. In the top panel (a), the total particle yield as a function of pseudorapidity $\eta$ is shown as a solid blue curve, along with the individual contributions of pions, kaons, and protons as red dashed, dot-dashed, and dotted lines. As can be inferred by the normalized residuals shown in the bottom panel, the model describes the pseudorapidity spectrum very well, with only the rightmost data point having an error exceeding $1\sigma$. In particular, allowing for a rapidity-dependent $D_Y$ produces a considerably better agreement with the data near the midrapidity dip compared to our previous results \cite{hgw24}, where, due to the assumption of a constant diagonal diffusion, the particle yield at $y = \eta = 0$ was found to be too low.
In Figs.\,5 (b) and (c),
the model predictions are compared with transverse-momentum data from ALICE~\cite{ALICE2018}, obtained by averaging over a $|\eta| \leq 0.8$ pseudorapidity slice. Fig.\,5 (b) shows the whole experimentally available transverse momentum range up to $p_\perp = 50$ GeV in a log-log scale, with the total particle yield and the single species contributions represented as in Fig.\,5 (a). The model prediction clearly deviates from the experimental data beyond $p_\perp = 4$ GeV -- partly due to the fact that the subsequent transverse region was not included in the $\chi^2$ minimization, but mainly because hard processes take over that are not covered in a nonequilibrium-statistical model.  A closer look at the fitted region is given in Fig.\,5\,(c), where the comparison of the model to data is shown for $p_\perp \leq 4$ GeV using a linear scale, in the upper part. The corresponding normalized residuals in the bottom panel indicate the deviations from the data. 

The data set for 2.76 TeV Pb-Pb collisions shown in Fig.\,\ref{fig6} is from the ATLAS experiment~\cite{ATLAS2015}, and contains two-dimensional measurements of charged-hadron production in the window  $0 \leq \eta \leq 2$ and $0.5 \leq p_\perp \leq 150$ GeV. In Fig.\,6\,(a) the calculated particle-number density function $\mathrm{d}^2N/\mathrm{d}\eta\mathrm{d}p_\perp$ is shown as a two-dimensional contour plot with the logarithmic color scale on the right, and the normalized residuals are represented as superimposed rectangular bins with the linear color scale on top. The lower figure 6\,(b) presents the same results, but plotted $p_\perp$-binwise against pseudorapidity, with each curve in $\eta$ obtained by averaging the calculated differential yield over a $p_\perp$ bin. The model gives an overall good description of the data, with the exception of the outer pseudorapidity region $1.5 \leq \eta \leq 2$, where the experimental measurements strongly deviate from the prediction, with a discrepancy of more than $2.5 \sigma$ for the two outest bins. The  {apparent increase} of the data  {around $\eta\simeq2$} in the low-$p_\perp$ bins cannot be accurately described with the assumption of a diagonal diffusivity, nor is it an effect of the Jacobian transformation from the $(h,y)$ coordinate space. A more complex model would probably be needed to address this feature, unless it is due to an underestimate of the systematic experimental error -- especially when comparing to the ALICE marginal pseudorapidity data~\cite{ALICE2013,ALICE2017} that show a much smoother trend at both, 2.76 and 5.02 TeV.

\subsection{Central $\sqrt{s_\text{NN}} = 5.02$ TeV Pb-Pb} 


The calculated marginal pseudorapidity distribution for central Pb-Pb collision at $\sqrt{s_\text{NN}} = 5.02$ TeV is compared to ALICE data~\cite{ALICE2017} covering the midrapidity window $-3.5 \leq \eta \leq 5$ in Fig.\,\ref{fig7}. As for the lower energy, the total hadron production $\mathrm{d}N/\mathrm{d}\eta$ is shown as a solid blue line, while the partial contributions of pions, kaons, and protons as red dashed, dot-dashed, and dotted lines in Fig.\,7\,(a). Figures\,7\,(b) and (c)  present the model prediction in the transverse direction, obtained  {by} averaging $\mathrm{d}^2N/\mathrm{d}\eta\mathrm{d}p_{\perp}$ over the $|\eta| \leq 0.8$ pseudorapidity slice, in comparison with ALICE data~\cite{ALICE2018}. Fig.\,7\,(b) shows the whole measured $p_\perp$ range plotted on a log-log scale, Fig.\,7\,(c) analyzes only the $p_\perp \leq 4$ GeV fitted region on a linear scale, and gives the associate normalized residuals.   
The spectrum is qualitatively similar to that at $2.76$ TeV collision energy, apart from the expected higher total number of produced particles. Due to the change in slope of the 5.02 TeV $\mathrm{d}N/\mathrm{d}\eta$ ALICE data around $|\eta| \approx 4$ and the lack of a measured joint spectrum outside the midrapidity region, it proved difficult to obtain a reasonable fit result for the longitudinal diffusivity. The $\chi^2$ minimization tried to compensate for these features in the data, resulting in an overfitted set of parameters predicting a high yield at intermediate-to-high rapidity. To address this problem that has already been encountered in Ref.\,\cite{hgw24}, the 2.76 TeV result for $D_\parallel \Delta t$ is extrapolated to the higher energy by scaling it with the beam rapidity, based on the earlier finding \cite{Wolschin2012} that the width of the marginal pseudorapidity distribution is in good approximation proportional to the logarithm of the respective center-of-mass energy. Despite having reduced the free parameters by fixing $D_\parallel \Delta t$, the model provides a satisfactory description of the experimentally measured spectrum in both the longitudinal and transverse directions. 
\subsection{Predictions for $\sqrt{s_\text{NN}} = 5.36$ TeV Pb-Pb}
\begin{figure*}
	\begin{center}
		\centering
		\includegraphics[width=\linewidth]{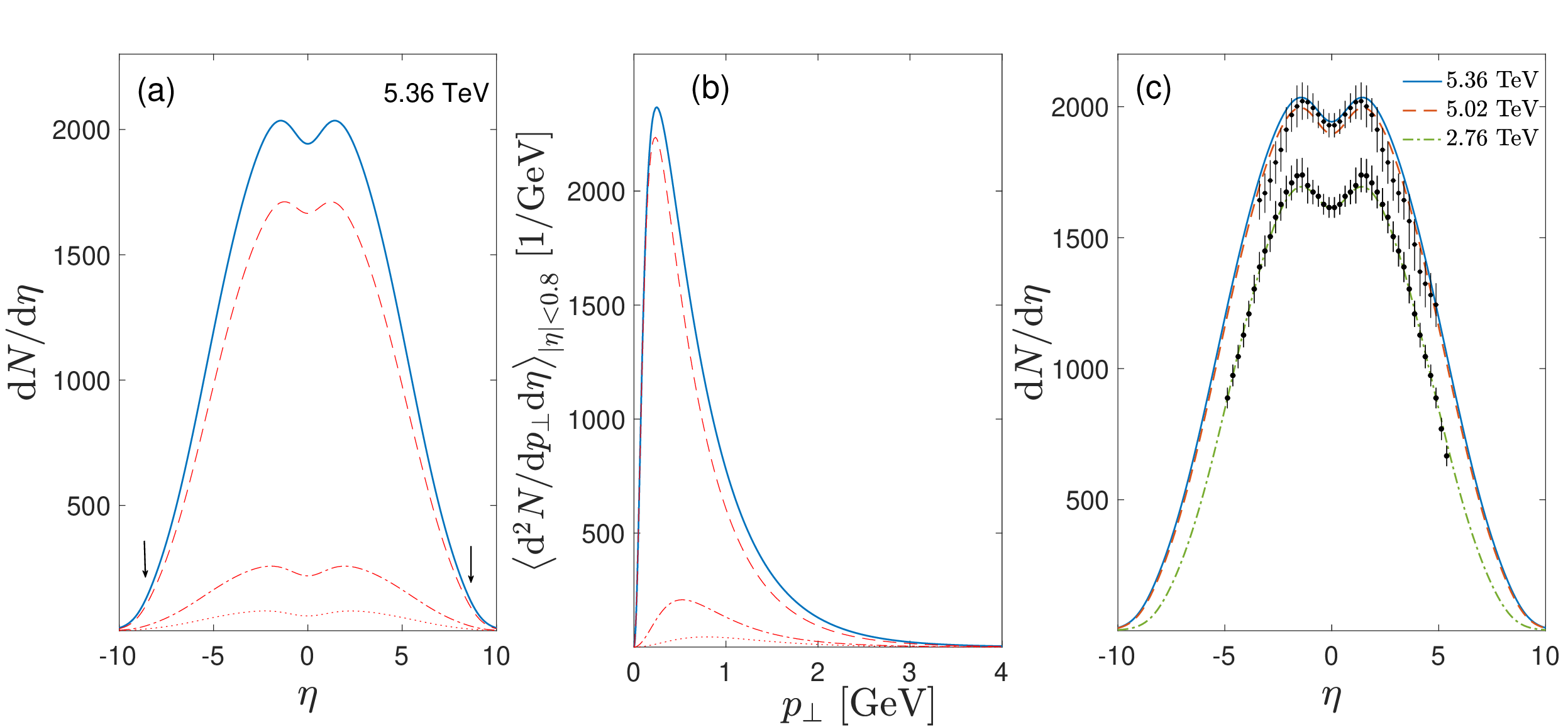} 
		\caption[Predicted marginal pseudorapidity spectrum for $\sqrt{s_\text{NN}} = 5.36$ TeV.]{(a) Predicted marginal pseudorapidity $\mathrm{d}N/\mathrm{d}\eta$ of produced charged hadrons in central $\sqrt{s_\text{NN}} = 5.36$ TeV Pb-Pb collisions. The curves are the total produced hadrons (solid) and the contribution of pions (dashed), kaons (dot-dashed), and (anti)protons (dotted). (b) Predicted joint spectrum $\mathrm{d}^2N/\mathrm{d}\eta\mathrm{d}p_\perp$ of produced charged hadrons in $\sqrt{s_\text{NN}} = 5.36$ TeV Pb-Pb central collisions, averaged over the $|\eta| < 0.8$ slice and plotted against $p_\perp$. (c) Predicted marginal pseudorapidity $\mathrm{d}N/\mathrm{d}\eta$ of produced charged hadrons in central $\sqrt{s_\text{NN}} = 5.36$ TeV Pb-Pb collisions (solid curve) compared to the result at $\sqrt{s_\text{NN}} = 5.02$ TeV (dashed) and $\sqrt{s_\text{NN}} = 2.76$ TeV (dot-dashed), with ALICE data as in Figs.\,5,7. }
		\label{fig8}
	\end{center}
\end{figure*}



During Run 3 at the LHC, the ALICE experiment has collected data of Pb-Pb collisions at $\sqrt{s_\text{NN}} = 5.36$ TeV in 2023. Here we use the relativistic diffusion model to make a prediction of the charged-hadron spectra at this energy. Since no data are yet available, we extrapolate the free parameters of the model at $\sqrt{s_\text{NN}} = 5.36$ TeV from the lower energies. Regarding the approach to thermalization in the longitudinal direction, the value of $D_Y \Delta t$ can be obtained by scaling it with the beam rapidity, as already done for $\sqrt{s_\text{NN}} = 5.02$ TeV. Since $\alpha$ reaches the lower boundary of $-1$ for both of the lower energies, this is assumed to be the case at 5.36 TeV as well. As for the parameters describing the transverse spectrum as shown in Table 1, it is more difficult to infer a scaling with the energy. Apart from the condition $c=5$ that is maintained, $\hat{T}$, $b$ and $d$ are also assumed to be proportional to $y_\text{b}$. 
For the total number of charged hadrons $N$, an investigation of the central and fragmentation sources in heavy-ion collisions at RHIC and LHC energies \cite{Wolschin2015} found a phenomenological scaling of the gluon-gluon total charged-hadron yield with the cube of the beam rapidity. We use this dependence to predict $N$ at 5.36 TeV. The extrapolated parameter set is given in Table~\ref{tab1}. The resulting spectra in the longitudinal and transverse directions are shown in Fig.\,\ref{fig8}, with (a) the predicted joint spectrum $\mathrm{d}^2N/\mathrm{d}\eta\mathrm{d}p_\perp$ of produced charged hadrons and (b) the corresponding transverse-momentum distribution.  According to Ref.\,\cite{Wolschin2015}, the particle yield at mid-rapidity is in good approximation 
\begin{equation}
	\left. \frac{\mathrm{d}N}{\mathrm{d}\eta} \right|_{\eta = 0} \approx 1150 \left(\frac{s_\text{NN}}{s_0}\right)^{0.165},
\end{equation} 
with $s_0$ = 1 TeV$^2$. The predictions of this phenomenological law agree with the existing data at both LHC energies, but slightly deviate from the present model prediction of the 5.36 TeV longitudinal spectrum. In Fig\,8\,(c), the marginal pseudorapidity hadron yields of the relativistic diffusion model as calculated using the spectral solution method are shown for all three energies in comparison with the presently available ALICE data. 
\section{Conclusions}
A spectral eigenfunction decomposition of the Fokker-Planck operator derived from a drift-diffusion model for partial thermalization of produced charged hadrons in relativistic heavy-ion collisions has been developed and implemented. The representation of the Fokker-Planck operator is based on a polynomial basis constructed with an orthonormality relation with respect to the stationary solution of the Fokker-Planck equation, which is the first eigenfunction of the associated differential operator. When considering the central fireball, which is the main source of particle production through gluon-gluon interactions at LHC energies, such a spectral method is successfully applied, and the convergence of the solution is shown to be very fast. It turns out to be difficult, however, to employ the same spectral representation to solve the FPE for the forward and backward fragmentation sources, because the overlap between initial and stationary distribution is very small.  

The high accuracy of the spectral solution method allows to compare the calculated spectrum in a large transverse-momentum region. The fit to experimental data from the ALICE and ATLAS collaborations has, however, been limited to $p_\perp < 4$ GeV, since at larger transverse momentum a perturbative QCD description is expected to account for the produced charged-hadron distribution from the non-equilibrium statistical approach used in this work. We have enhanced the relativistic drift-diffusion model of Ref.\,\cite{hgw24} using a rapidity-dependent diffusion, which produces a better description of the longitudinal spectra in the mid-rapidity region. The total reduced $\chi^2$ indicates significantly improved model results at both $\sqrt{s_\text{NN}} = 2.76$ and 5.02 TeV. Extrapolating
the model parameters to $\sqrt{s_\text{NN}} = 5.36$ TeV, we have also predicted the charged-hadron spectra for Run 3 at the LHC in the longitudinal and transverse directions.

Future developments of the model should include a microscopic derivation of the rapidity dependence of the diffusion coefficient, and possibly consider a transverse dependence as well. Including off-diagonal components in the diffusion matrix could also provide a better physical representation of charged-hadron production, but is certainly challenging to implement in the present solution scheme. 

An extension to non-central collisions would require to consider the third spatial dimension, which would be problematic to tackle with spectral methods. 
The inclusion of the fragmentation sources is essential to describe baryon stopping, and could possibly be addressed in the spectral solution scheme by employing a different representation of the Fokker-Planck operator. 
	
\begin{acknowledgement}
We thank Johannes H\"olck for discussions and remarks.
\end{acknowledgement}
\section*{Declarations}
\begin{itemize}
\item Availability of data and materials:
The experimental data that are compared with our model calculations are available on https://www.hepdata.net.
The model results shown in the figures
 are available from the authors upon reasonable request.
\item Code availability: 
The Matlab-codes can be made available upon request.
\item Authors' contributions:
AR: Adapting the spectral eigenfunction method to the relativistic diffusion model, numerical solutions, plots; Writing - parts of the draft. GW: Concept, Methodology; Writing - parts of the draft; supervision, review and editing.

\end{itemize}

%
\end{document}